\documentclass[aps,prl,superscriptaddress,twocolumn,amsmath,amssymb]{revtex4}

\usepackage{amsmath}
\usepackage{graphicx}
\usepackage{color}
\usepackage{subfigure}

\newcommand{\sci}{Science }

\begin{document}

\title{Adiabatic Preparation of a Heisenberg Antiferromagnet Using an Optical Superlattice}

\author{Michael Lubasch}
\affiliation{Max-Planck-Institut f\"ur Quantenoptik, Hans-Kopfermann-Stra\ss{}e 1, 85748 Garching, Germany.}
\author{Valentin Murg}
\affiliation{Fakult\"at f\"ur Physik, Universit\"at Wien, Boltzmanngasse 3, 1090 Wien, Austria.}
\author{Ulrich Schneider}
\affiliation{Fakult\"at f\"ur Physik, Ludwig-Maximilians-Universit\"at M\"unchen, Schellingstra\ss{}e 4, 80799 M\"unchen, Germany.}
\author{J.~Ignacio Cirac}
\affiliation{Max-Planck-Institut f\"ur Quantenoptik, Hans-Kopfermann-Stra\ss{}e 1, 85748 Garching, Germany.}
\author{Mari-Carmen Ba\~nuls}
\affiliation{Max-Planck-Institut f\"ur Quantenoptik, Hans-Kopfermann-Stra\ss{}e 1, 85748 Garching, Germany.}

\date{\today}

\begin{abstract}
We analyze the possibility to prepare a Heisenberg antiferromagnet with cold fermions in optical lattices, starting from a band insulator and adiabatically changing the lattice potential.
The numerical simulation of the dynamics in 1D allows us to identify the conditions for success, and to study the influence that the presence of holes in the initial state may have on the protocol.
We also extend our results to two-dimensional systems.
\end{abstract}

\maketitle

Ultracold atoms trapped in optical lattices offer a unique possibility to experimentally explore strongly correlated states of quantum matter.
Currently, one of the main experimental challenges in this field is the preparation of a Heisenberg antiferromagnet (AFM),
which represents the necessary next experimental step towards a true quantum simulator of the fermionic Hubbard model~\cite{Esslinger}.

Although the creation of a fermionic Mott insulator (MI) has recently been reported~\cite{Joerdens1, Schneider1}, the realization of antiferromagnetic order requires temperature and entropy significantly lower than presently achieved~\cite{Werner, Joerdens2}, despite many existing proposals for direct cooling within the lattice~\cite{McKay}.
An alternative to the direct generation is to use an adiabatic protocol~\cite{Ho, Sorensen}.
In such a scheme, it is desirable to tune interactions initially to give a ground state with very low entropy.
Then, they are changed slowly, until the Heisenberg Hamiltonian is realized at the end.
If the process is adiabatic, the entropy will stay low and the final state will be the desired AFM.
The following questions immediately arise: What are the conditions to achieve adiabaticity?
What occurs if these conditions are too restrictive and cannot be met?
And, how will the protocol be affected by a finite temperature and the presence of a harmonic trap?

In this Letter, we propose a specific adiabatic scheme and analyze these issues.
Our adiabatic protocol is the first to attain an AFM with ultracold fermions within feasible timescales, even in the presence of experimental imperfections.
Additionally, we show that it is possible to realize antiferromagnetic order on a part of the sample in a shorter time than required for the whole system.
Finally, we simulate the dynamics of holes to demonstrate their destructive effect on the AFM and devise a strategy to control them.

The initial ground state of our protocol is a band insulator (BI), which is transformed first to an array of decoupled singlets and finally to the AFM, by adiabatically changing the depth of two superimposed optical lattices.
A BI is easier to prepare with low entropy than a MI for two reasons.
On the one hand, its energy gap is given by the band gap, which is much larger than the interaction energy (MI gap) and favors a redistribution of the entropy towards the surrounding metallic shell~\cite{Schneider2}.
On the other hand, the preparation can be done using weakly or noninteracting atoms, thereby avoiding the long timescales associated with mass and entropy transport at higher interactions~\cite{Schneider3}.

For the one-dimensional case, we simulate the fermionic \mbox{$t-J$} model with matrix product states (MPS)~\cite{MPS}.
We first identify the adiabatic conditions that allow the preparation of the antiferromagnetic state in an ideal case with no defects.
Second, we study how these conditions are relaxed if one imposes that antiferromagnetic order is only obtained on a subset of fermions around the center of the sample.
We observe that, when restricted to a middle sublattice, adiabaticity is determined by an effective gap related to this sublattice and not by the gap of the total system.
Third, we include the presence of holes in the initial state, expected to occur in real experiments due to the finite temperature.
The large initial energy of the holes can in principle destroy the AFM as they delocalize inside the sample.
We find that, if the holes are initially located at the outer part of the sample, as expected in an experiment, a tradeoff can be reached between the degree of adiabaticity of the process and the distance the holes travel inside the chain, so that the antiferromagnetic order is still produced in the center.
Moreover, we show that a harmonic trap can prevent the destructive effect of holes by confining them to the outside of the sample.
Finally, via projected entangled pair states (PEPS)~\cite{PEPS1, PEPS2}, we complement our analysis with a simulation of the two-dimensional \mbox{$t-J$} model of hardcore bosons with antiferromagnetic interaction.
This setting is easier to investigate numerically than the corresponding 2D fermionic system and provides evidence that the physics studied in the one-dimensional case can be extrapolated to understand the conditions of an equivalent scheme in 2D.

\begin{figure}[!t]
\centering
\includegraphics[width=0.45\textwidth]{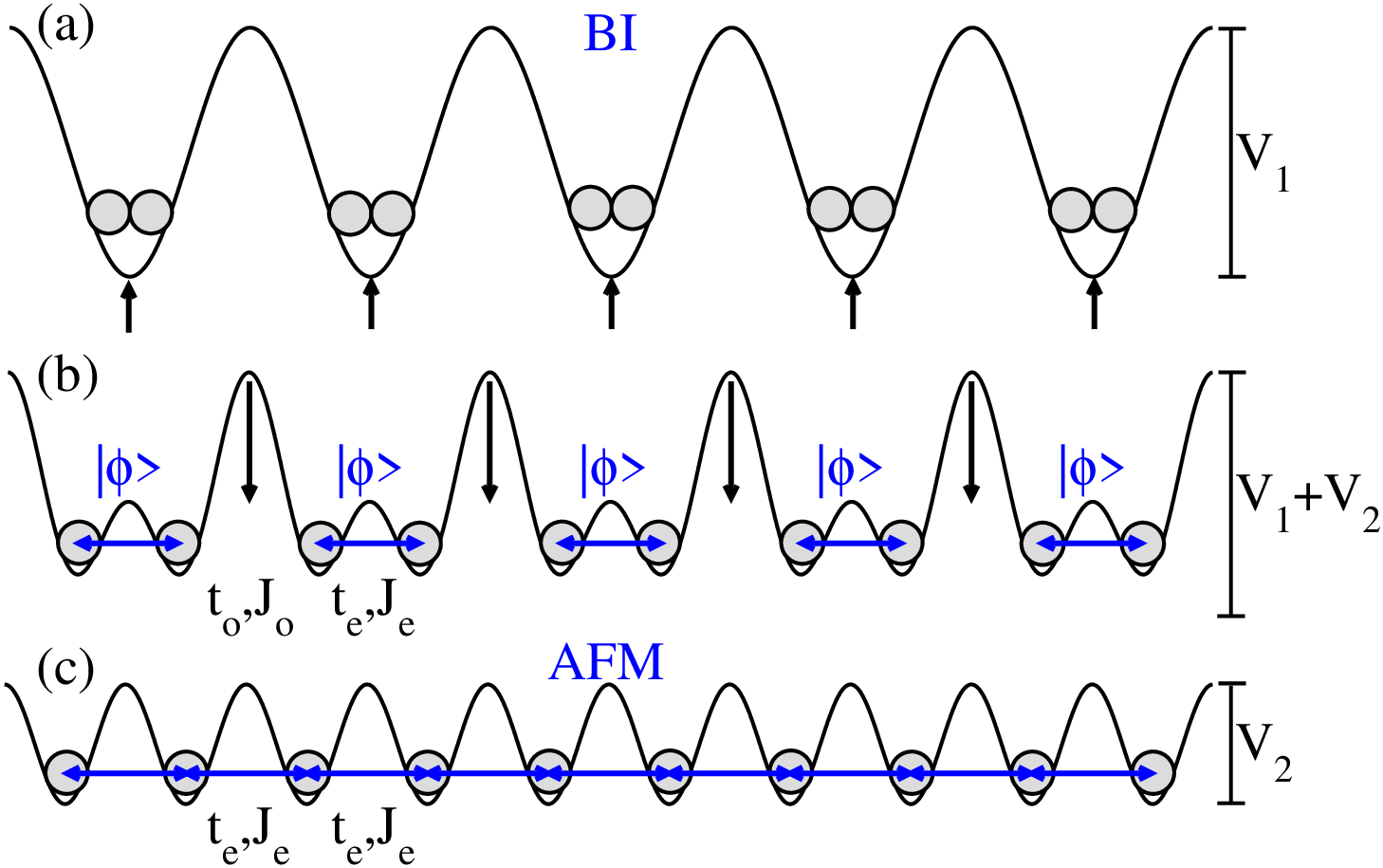}
\caption{\label{fig:1}The proposed adiabatic protocol.
First, a BI in a lattice with depth $V_1$ (a) is transformed to a product of singlets \mbox{$|\phi\rangle$} (b) by slowly switching on a second lattice with depth $V_2$ and half the original wave length.
Then, by lowering the barrier $V_{1} \to 0$, the system turns into an AFM (c).}
\end{figure}

In the following, unless stated otherwise, we will focus on the one-dimensional case and consider a two component Fermi gas in an optical lattice.
The physical setting consists of two adiabatic stages, depicted in Fig.~\ref{fig:1}.
The first transition has already been realized experimentally~\cite{Sebby-Strabley, Trotzky1, Trotzky2} and can be straightforwardly described, so that we can focus on the second one and take Fig.~\ref{fig:1}(b) as the initial state for our theoretical study.
In this situation, the system is governed by a one-band \mbox{$t-J$} model.
This model emerges in the limit of strong interactions from the Hubbard model, which describes ultracold atoms in optical lattices~\cite{Jaksch}.
We consider a bipartite \mbox{$t-J$} Hamiltonian with different couplings for even and odd links, \mbox{$H=H_{e}+H_{o}$},
where
\begin{eqnarray}
H_{\ell}&=&-t_{\ell} \sum_{k\in\ell,\,\sigma=\uparrow,\downarrow}(c_{k, \sigma}^{\dagger}c_{k+1, \sigma}+\mathrm{H.c.})
 \\
&&+J_{\ell} \sum_{k\in\ell} \left( \mathbf{S}_{k}\mathbf{S}_{k+1}-\frac{n_{k}n_{k+1}}{4}\right),\quad \ell=e, o.
\label{eq:1}
\nonumber
\end{eqnarray}
The superexchange interaction $J_{\ell}$ and the tunneling parameter $t_{\ell}$ are related through the on-site interaction $U$ as \mbox{$J_{\ell}=4 t_{\ell}^2/U$}.
We fix the couplings on the even links, \mbox{$t_{e}=t$} and $J_{e}=J$, and choose a linear ramping of the superexchange interaction on the odd links, over total ramping time $T$ so that
 \mbox{$J_{o}(\tau)=J \cdot \tau / T$} and 
 \mbox{$t_{o}(\tau)=t \cdot \sqrt{\tau / T}$},
for \mbox{$0 \le \tau \le T$}~\cite{Ramping}.
In the following, we set \mbox{$J=1$}.
A harmonic trap is included by adding a term \mbox{$V_{\mathrm{t}}\sum_k(k-k_0)^2 n_k$} to Eq.~(1).

The ramping time from Fig.~\ref{fig:1}(b) to \ref{fig:1}(c) needs to be long enough such that the final state is close to the true AFM.
The required time to ensure a certain degree of adiabaticity can be seen to scale as \mbox{$T \propto 1/\Delta^2$}~\cite{Zener}, where $\Delta$ is the minimum gap between the ground and the first excited state during the evolution.
A closer look at the relevant energy levels reveals that in this adiabatic transition the gap decreases monotonically from $J$ to the Heisenberg gap, which vanishes in the thermodynamic limit, and there is no phase transition occurring in between~\cite{Matsumoto}.

The adiabaticity of the evolution and the final antiferromagnetic order can be probed by two experimentally accessible observables.
The first one, the squared staggered magnetization,
\mbox{$M_{\mathrm{stag}}^{2} = \frac{1}{N^{2}} \sum_{l,m=1}^{N} (-1)^{l+m} \langle \mathbf{S}_{l} \mathbf{S}_{m} \rangle$},
is the antiferromagnetic order parameter and can be determined by noise correlations~\cite{Bruun}.
The second one is the double well singlet fraction,
\mbox{$P_0=\frac{2}{N}\sum_{k \in e}\left(\frac{1}{4}-\langle\mathbf{S}_{k}\mathbf{S}_{k+1}\rangle\right)$}.
Since the initial state, Fig.~\ref{fig:1}(b), has a pure singlet in each double well, measuring \mbox{$P_0 \neq 1$} at the end indicates a change in the state.
The generation and detection of singlet and triplet dimers in double well lattices has been recently reported~\cite{Trotzky2}.
Whereas the squared staggered magnetization is experimentally detected over the whole sample and captures information on long-range correlations, the singlet fraction can be determined \emph{in situ} and hence allows us to probe parts of the sample.
For the Heisenberg antiferromagnetic chain in the thermodynamic limit, these observables take on the values $M_{\mathrm{stag}, \mathrm{TD}}^{2}=0$ and $P_{0, \mathrm{TD}} \approx 0.693$~\cite{Hulthen}, while in 2D \mbox{$M_{\mathrm{stag}, \mathrm{TD}}^{2} \approx 0.0945$} and \mbox{$P_{0, \mathrm{TD}} \approx 0.585$}~\cite{Sandvik}.
Notice that for the finite 1D systems considered in this work, $M_{\mathrm{stag}}^{2}$ does not vanish, but has a sizable value, comparable to the 2D thermodynamic limit~\cite{Supplementary}.

\paragraph{Absence of holes.}
Using the numerical simulation of the chain dynamics with MPS, we investigate the state at the end of the protocol for varying ramping time.
To characterize the antiferromagnetic order independently of the system size, we define the relative quantities
\mbox{$m^{2}(T):=M_{\mathrm{stag}}^{2}(T)/M_{\mathrm{stag}, \mathrm{AFM}}^{2}$}, \mbox{$p_{0}(T):=P_{0}(T)/P_{0, \mathrm{AFM}}$}, and
\mbox{$e_{\mathrm{spin}}(T):=E_{\mathrm{spin}}(T)/E_{\mathrm{spin}, \mathrm{AFM}}$}, where the denominator is the expectation value of the observable in the true AFM for a given lattice~\cite{Supplementary}.
For the last quantity, $E_{\mathrm{spin}}$ is the expectation value of the spin term in the total Hamiltonian of Eq.~(1).

\begin{figure}[h]
\centering
\includegraphics[width=0.48\textwidth]{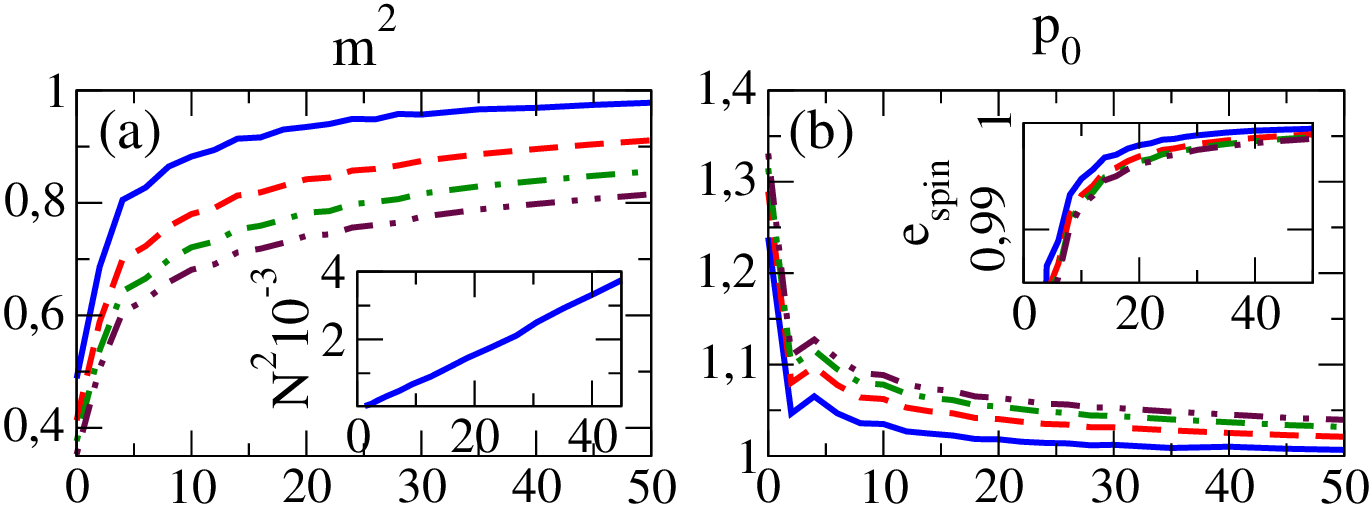}
\includegraphics[width=0.48\textwidth]{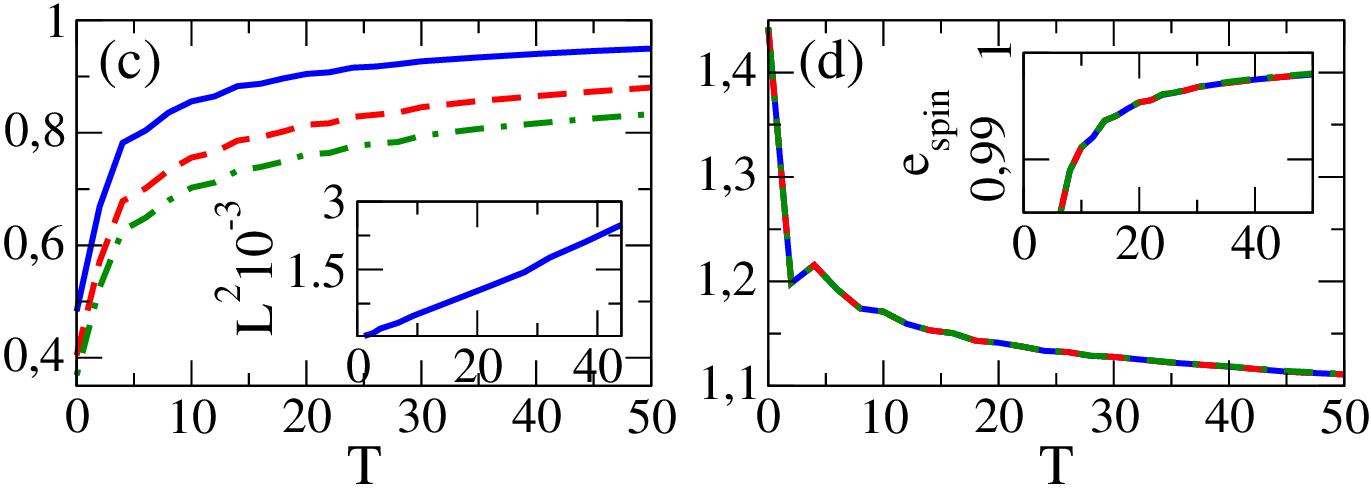}
\caption{\label{fig:2}Absence of holes.
(a), (b) $m^{2}$, $p_{0}$, and $e_{\mathrm{spin}}$ as functions of the ramping time $T$, for chain length \mbox{$N=22$} (solid blue lines), \mbox{$42$} (dashed red lines), \mbox{$62$} (dash-dotted green lines), and \mbox{$82$} (dash double-dotted brown lines).
The inset of (a) shows the squared size $N^2$ of the longest chain reaching a fixed \mbox{$m^{2} = 0.85$} at ramping time $T$, and reveals the scaling \mbox{$T \propto N^{2}$}.
(c), (d) Same quantities as above, evaluated on sublattices of length \mbox{$L=22$} (solid blue lines), \mbox{$42$} (dashed red lines), and \mbox{$62$} (dash-dotted green lines) for \mbox{$N=82$}.
Now, the inset of (c) shows the squared size $L^2$ of the largest sublattice reaching \mbox{$m^{2} = 0.85$} at ramping time $T$, and reveals the scaling \mbox{$T \propto L^{2}$}.
All results were obtained with MPS of bond dimension \mbox{$D=60$} and Trotter step \mbox{$\delta t = 0.02$}~\cite{Bond}.}
\end{figure}

Figure~\ref{fig:2} shows our results for an ideal case with no holes in the initial state, with all relative quantities converging to $1$, as expected, in the limit of large $T$ [Figs.~\ref{fig:2}(a),(b)].
The ramping time necessary to reach a certain relative magnetization $m^2$ grows with the system size.
If we study, given $T$, which is the largest system for which a fixed value $m^{2}$ can be achieved [Fig.~\ref{fig:2}(a) inset], we find \mbox{$N^{2} \propto T$}, which is consistent with the adiabaticity condition for a gap closing like \mbox{$\Delta \propto 1/N$}~\cite{Affleck}.

For very long chains, the required $T$ might not be experimentally feasible.
Remarkably, this does not exclude the preparation of antiferromagnetic order on large systems.
We may evaluate the magnetization over a sublattice in the center of the sample.
If, given $T$ and $N$, we ask for the largest sublattice size $L$ for which the magnetization reaches a fixed value [Fig.~\ref{fig:2}(c)], we find a scaling \mbox{$T \propto L^{2}$}, as governed by an effective local gap, and not by the gap of the total system.
In contrast to $m^{2}$, the observables $p_{0}$ and $e_{\mathrm{spin}}$ do not depend on $L$ [Fig.~\ref{fig:2}(d)].
This can be understood from the fact that $p_{0}$ and $e_{\mathrm{spin}}$ are determined by a two-site observable \mbox{$\mathbf{S}_{k}\mathbf{S}_{k+1}$} averaged over a sublattice of length $L$, whereas $m^{2}$ is a true $L$-site observable,
and thus effectively probes the adiabaticity on the sublattice.
These observations can be addressed analytically by means of spin wave theory~\cite{InPreparation}.
The experimental consequence is that with a large sample, high values of $m^2$ can be obtained in short ramping times $T$ on small parts of the system, $L \propto \sqrt{T}$.

\begin{figure}[h]
\centering
\includegraphics[width=0.48\textwidth]{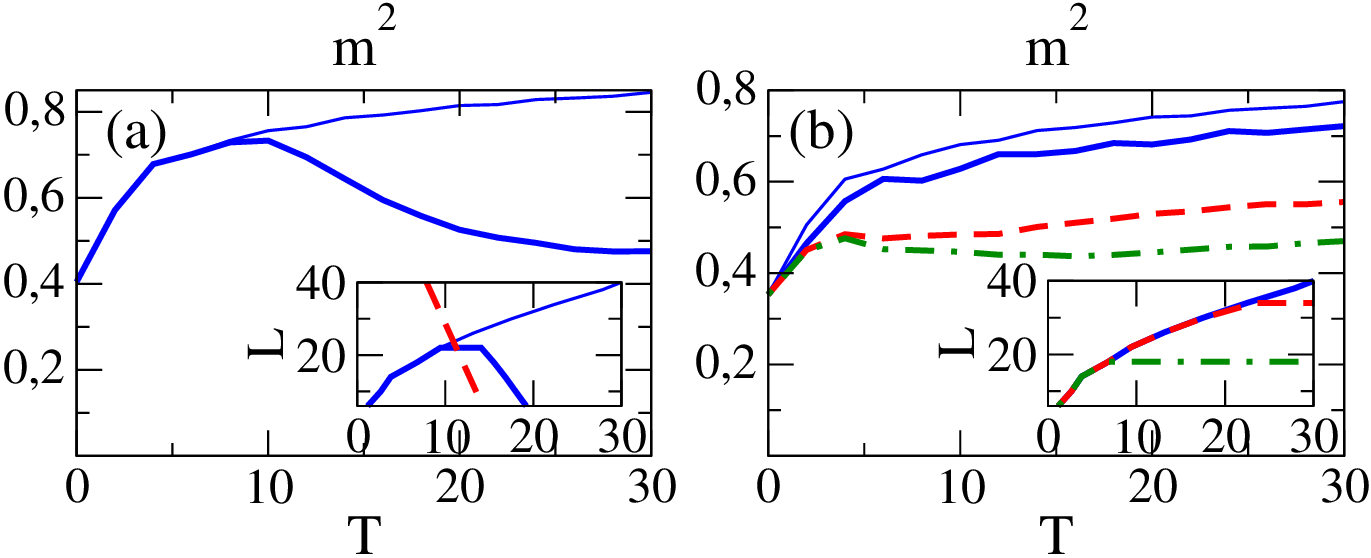}
\caption{\label{fig:3}Effect of holes and harmonic trap.
(a) $m^{2}$ as a function of the ramping time $T$, evaluated on a sublattice of length \mbox{$L=42$} for \mbox{$N=86$}, without holes (thin solid line), and with initially 2 holes at each boundary (thick solid line), and \mbox{$t=2$}.
The inset shows the largest sublattice size $L$ reaching \mbox{$m^{2} = 0.85$} at ramping time $T$, for the two cases of the main plot, and the size of the hole-free region (dashed red line).
(b) $m^{2}$ evaluated on a middle sublattice of length \mbox{$L=82$} for \mbox{$N=102$}, with no holes (thin solid line), and with initially 10 holes at each boundary and a harmonic trap of strength \mbox{$V_{\mathrm{t}}=0.004$} (dash-dotted green line), $0.006$ (dashed red line), and $0.02$ (solid blue line), and \mbox{$t=3$}.
For $V_{\mathrm{t}}=0.2$ (not shown) the exact behavior of the ideal case is recovered.
Again, the inset shows the largest sublattice size $L$ reaching \mbox{$m^{2} = 0.85$} at ramping time $T$, for the cases of the main plot.
All results were obtained with MPS of bond dimension \mbox{$D=60$} and Trotter step \mbox{$\delta t = 0.02$}~\cite{Bond}.}
\end{figure}

\paragraph{Effect of holes.}
In a real experiment, the finite temperature causes the sample to be in a thermal mixture.
As a consequence, localized holes will be present in the initial state, Fig.~\ref{fig:1}(b).
Since the double wells are decoupled, the wave function of a hole will be an equal superposition of being in the left and in the right side of a single double well.
Our simulation reveals that holes have a highly destructive effect on magnetic order.
As seen in Fig.~\ref{fig:3}(a), a few holes initially located on the boundary of the sample are enough to cause a dramatic reduction of the final staggered magnetization.

We observe that the dynamics of holes can be qualitatively well understood using a simplified picture, in which the spreading of an initially localized hole, propagating in an antiferromagnetic background, is modeled by a free particle.
This picture is accurate in the limit \mbox{$t \gg J$}, when the spin term in Eq.~(1) is negligible.
We checked that it is also valid in the whole experimentally reasonable parameter regime by comparing the behavior of single holes to that of free particles with the same initial wave function~\cite{InPreparation}: for the relevant range of times and tunneling values, \mbox{$2 \leq t \leq 8$}, the hole spreads like a free particle with a maximal velocity $v=2 t$.
In the case of small \mbox{$t = 2$}, a hole with higher initial energy causes a higher spin excitation, while for a large \mbox{$t = 8$}, the hole excites the spin background by \mbox{$\Delta E_{\mathrm{spin}} \approx 0.5$} in the beginning, independent of its initial kinetic energy.
This can be understood by assuming a simple classical N\'eel background, where the delocalization of a hole, initially positioned at a boundary, breaks up exactly one antiferromagnetic bond.
This assumption should become a good approximation for the hole dynamics in the regime \mbox{$t \gg J$}, where the timescale of the delocalization is much faster than the timescale of the reacting spin background.
In all cases, the squared staggered magnetization is reduced substantially during propagation of the hole until it reaches a minimum after the hole has travelled once over the whole sample.
We found that the magnetization reduction depends only weakly on the initial kinetic energy, and on $t$, but it depends strongly on the number of holes in the sample.

The simplified free particle picture allows us to interpret the results from Fig.~\ref{fig:3}(a).
In particular, the strong magnetization drop around \mbox{$T \approx 8$} indicates the arrival of holes at the middle sublattice.
We can roughly predict this arrival time from the spreading of a free particle wave function, which, after time $T$, will have covered \mbox{$\frac{4}{3} t \cdot T$} sites, taking the ramping of the lattice into account.
We observe that the magnetization for a given sublattice $L$ behaves like in Fig.~\ref{fig:2}(c) only for short ramping times, while the region is hole-free, until the holes reach the sublattice [Fig.~\ref{fig:3}(a), inset].

\paragraph{Harmonic trap.}
The simplified picture described above points out that the negative effect of holes can be controlled by the presence of a trap.
An external potential changes sign for a hole and effectively turns into an inverse trap, capable of confining the holes to the outer parts of the chain.
The trap strength should be chosen as large as possible without exceeding the on-site interaction $U$, what would destroy the MI Fig.~\ref{fig:1}(c).

The results of the dynamics within the harmonic trap are shown in Fig.~\ref{fig:3}(b).
From energy considerations, a hole delocalizes at most by $\pm 2t$~\cite{Supplementary}.
As the trap strength is increased, the holes get more localized on the outside.
As a consequence, the magnetization of the total sample increases, and the behavior of the hole-free case is recovered.

\paragraph{Two-dimensional case.}
For the 2D case, the adiabatic setting consists of an array of initially decoupled chains like Fig.~\ref{fig:1}(b), connected by a transverse lattice with the time-dependent couplings $J_{o}$ and $t_{o}$.
Different to the 1D case, this system exhibits a phase transition in the thermodynamic limit at \mbox{$J_{o} \approx 0.5$}~\cite{Matsumoto}.

Although the numerical simulation of the 2D setting is much more demanding than the one for chains, we can relatively easily obtain results for hardcore bosons on lattices of moderate size.
In the absence of holes, the bosonic and fermionic \mbox{$t-J$} models are equivalent, and our simulations serve to test our protocol on a 2D system.
Similarities between both models under inclusion of holes are a subject of current research~\cite{Boninsegni}, but our simulations can still provide a qualitative indication of the controlling effect of the trap.
It is worth noticing that the AFM can also be realized with ultracold bosons~\cite{BosonicAFM}.

\begin{figure}[h]
\centering
\includegraphics[width=0.48\textwidth]{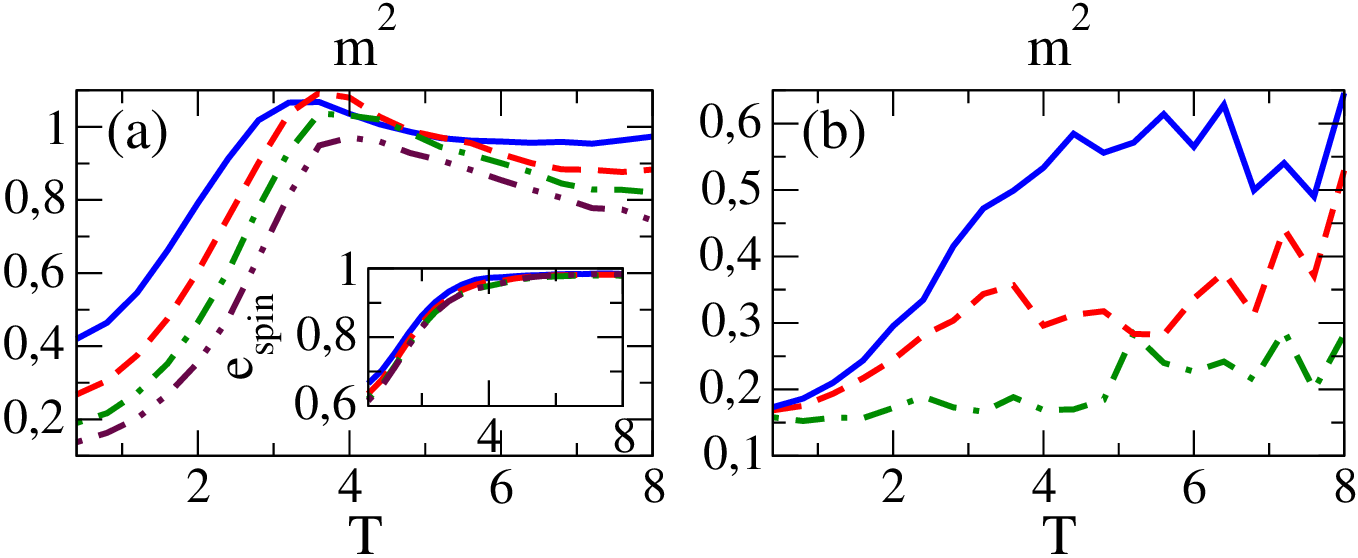}
\caption{\label{fig:4}Performance in 2D.
(a) $m^{2}$ and $e_{\mathrm{spin}}$ as functions of the ramping time $T$, for systems of size \mbox{$N = 4 \times 4$} (solid blue line), \mbox{$6 \times 6$} (dashed red line), \mbox{$8 \times 8$} (dash-dotted green line), and \mbox{$10 \times 10$} (dash double-dotted brown line).
(b) $m^{2}$ for \mbox{$N = 8 \times 8$}, with initially 4 holes at the boundary and no harmonic trap (dash-dotted green line), and with a trap of strength \mbox{$V_{\mathrm{t}}=0.25$} (dashed red line), and $2.5$ (solid blue line), and \mbox{$t=2.5$}.
The results were obtained with PEPS of bond dimension \mbox{$D=4$} (a) and \mbox{$D=2$} (b) and Trotter step \mbox{$\delta t = 0.03$}~\cite{Bond}.}
\end{figure}

In the ideal case of no holes, we observe [Fig.~\ref{fig:4}(a)] that, whereas the energy converges quickly, the magnetization does not, and thus we cannot claim convergence to the true AFM within the numerically accessible ramping times studied here.
Remarkably enough, we find that for a $10\times10$ lattice, a significant magnetization value $m^2$ is obtained in times of the same order of magnitude as for a chain of length $10$, suggesting that the generation of antiferromagnetic order on much larger 2D lattices will be experimentally possible within reasonable timescales.
Upon hole injection, a similarly dramatic magnetization reduction is observed [Fig.~\ref{fig:4}(b)] which can be controlled by the presence of a harmonic trap, as in the 1D case.

\paragraph{Discussion.}
We have proposed and analyzed an adiabatic protocol, suitable to prepare an antiferromagnetically ordered state in an optical lattice, even from an initial state containing defects.
The timescales for the finite systems studied in this work lie well within the range of current experiments.
Furthermore, we have observed that antiferromagnetic order can be produced in a sublattice in times governed only by its size.

This scheme offers several advantages over other proposals.
First, starting from a BI simplifies the preparation of a sufficiently low entropy initial state.
Additionally, the initial ground state in Fig.~\ref{fig:1}(b) already features the final $\mathrm{SU(2)}$ symmetry of the AFM, so that the number of excited states to which the evolution couples is minimum, as compared with an alternative proposal~\cite{Sorensen} with only $\mathrm{U(1)}$ initial symmetry.
Moreover, since hole doping is experimentally feasible, the same procedure can possibly be used to prepare the ground state with varying hole densities.
This would open the door to the experimental exploration of open questions in condensed matter theory, ultimately the existence of $d$-wave superconductivity in the \mbox{$t-J$} model.

\paragraph{Acknowledgements.}
We are grateful to Immanuel Bloch for inspiring discussions.
This work was partially funded by the European projects Quevadis and AQUTE.

\clearpage
\onecolumngrid
\appendix

\section{Supplementary Material}

In this supplementary material, we describe the numerical method used in our article, provide absolute values of experimental observables and discuss the numerical errors.

\section{Numerical method}

We employ Matrix Product States (MPS) in 1D and Projected Entangled Pair States (PEPS) in 2D to compute ground states and time evolution~\cite{WhiteDMRGSupp,DMRGSupp,MPS1Supp,MPS2Supp,PEPS1Supp,PEPS2Supp}. For a chain of $N$ quantum systems, each having physical dimension $d$, a MPS has the form
\begin{eqnarray*}
 |\psi_{MPS}\rangle & = & \sum_{s_{1}, s_{2}, \ldots, s_{N}} \mathrm{tr}(A[1]^{s_{1}} A[2]^{s_{2}} \ldots A[N]^{s_{N}}) |s_{1} s_{2} \ldots s_{N}\rangle \qquad ,
\end{eqnarray*}
where the $s_{i}$ run from $1$ to $d$ and the $A[i]^{s_{i}}$, for a fixed $s_{i}$, are matrices of dimension \mbox{$D \times D$}.
The number of variational parameters in the ansatz is determined by the bond dimension, $D$, which also bounds the entanglement of the MPS.
The MPS family constitutes a very good approximation to ground states of gapped local Hamiltonians in one dimension~\cite{GSmpsSupp},
and it has become a very successful tool for the study of quantum many-body systems.
PEPS constitute the natural generalization of the MPS ansatz to larger dimensions~\cite{PEPS1Supp,PEPS2Supp}, and they are also known to be good approximations for ground and thermal states of gapped local Hamiltonians~\cite{GSpepsSupp}.

The basic algorithms for the study of many-body systems using these families of states are the variational search for the ground state and the simulation of time evolution.
A detailed description of these algorithms, which are similar for 1D and 2D systems,
can be found in the review paper~\cite{PEPS3Supp} (see also references therein).
The computational cost of MPS algorithms scales as \mbox{$\mathcal{O}(ND^{3})$}, which allows the simulation of large chains and bond dimensions.
PEPS algorithms scale as \mbox{$\mathcal{O}(ND^{10})$}, so that 2D simulations are much more demanding, and only smaller values of the virtual bond dimension $D$ can be considered.

In the variational search for ground states, the numerical error comes from limiting the bond dimension to 
a certain maximum value.
Comparing the results with those obtained after running the search with a larger $D$ gives an estimation 
of the magnitude of this truncation error.
Typically, the algorithm is run repeatedly with increasing bond dimension until convergence is
achieved within the desired numerical precision.

The simulation of time evolution of MPS (or PEPS) is based on a Suzuki-Trotter decomposition of the evolution operator~\cite{DMRGSupp}.
This introduces another source of numerical error, in addition to the truncation of the bond dimension. The magnitude of
this Trotter error can be controlled by decreasing the size of the time step, $\delta t$, or using a higher order decomposition
of the exponential.
In particular, we use a second order Trotter decomposition in the case of one-dimensional simulations, while a first order decomposition is used in 2D.
For each time step, the evolution operator for $\delta t$ is applied
and an optimal MPS or PEPS approximation to the evolved state is found as described in~\cite{MPS2Supp}.
The magnitude of the Trotter error is controlled by comparing results for various values of $\delta t$,
and the truncation error, as in the ground state search, is estimated from the comparison of results for
different $D$.

\section{Absolute values of experimental observables}

The relative quantities shown in the main text result from normalizing the computed expectation values at the end of the evolution to the corresponding values in the true AFM ground state. For completeness, we show in this section the actual absolute values obtained for each observable, as well as the reference AFM values, and discuss the convergence of the numerical results.

As explained in the article, we model our system by a bipartite \mbox{$t-J$} Hamiltonian, that, in 1D, reads:
\begin{eqnarray}\label{eq:tJSupp}
 H \! = \! - \! t_{\mathrm{e}} \! \sum_{k \in \mathrm{e},\,\sigma} \!
                               ( c_{k, \sigma}^{\dag} c_{k+1, \sigma} + \mathrm{h.c.} ) \!
           + \! J_{\mathrm{e}} \! \sum_{k \in \mathrm{e}} \! \Big( \mathbf{S}_{k} \mathbf{S}_{k+1} -
                                                            \frac{n_{k} n_{k+1}}{4} \Big) \!
           - \! t_{\mathrm{o}} \! \sum_{k \in \mathrm{o},\,\sigma} \!
                               ( c_{k, \sigma}^{\dag} c_{k+1, \sigma} + \mathrm{h.c.} ) \!
           + \! J_{\mathrm{o}} \! \sum_{k \in \mathrm{o}} \! \Big( \mathbf{S}_{k} \mathbf{S}_{k+1} -
                                                            \frac{n_{k} n_{k+1}}{4} \Big) \, ,
\end{eqnarray}
where the subscripts e and o denote even and odd sites and double occupancy of sites is forbidden, as implicitly assumed for the \mbox{$t-J$} model.
The couplings on even links are constant, $t_{\mathrm{e}}=t$ and $J_{\mathrm{e}}=J$,
while the time-dependent odd couplings are increased from $0$ to their final values, $t$ and $J$, during a total ramping time $T$,
according to \mbox{$t_{\mathrm{o}}(\tau) = t \cdot \sqrt{\tau / T}$} and \mbox{$J_{\mathrm{o}}(\tau) = J \cdot \tau / T$}.
We set \mbox{$J=1$}.
The two-dimensional system consists of several such chains, coupled in the transverse direction by $t_{\mathrm{o}}(\tau)$, $J_{\mathrm{o}}(\tau)$.

\subsection{Reference values in the AFM ground state}

We use the algorithms described above to compute numerically the true AFM ground state for various lattices, 
as shown in Tab.~\ref{tab:1Supp} (for chains of lengths $N=22-82$).
The numerical convergence is checked by comparing the values obtained using bond dimensions $D=80$ and $100$.
As can be seen from the values in the table, the maximum relative error is of the order $10^{-5}$ for the magnetization.
Note that the energy is correct up to $10^{-9}$.
We use these values as the reference for adiabaticity of the total chain of length $N$.

\begin{table}
 \begin{tabular}{ c || c | c | c | c | c | c }
  N & $M_{\mathrm{stag}}^{2}(D=80)$ & $M_{\mathrm{stag}}^{2}(D=100)$ & $P_{0}(D=80)$ & $P_{0}(D=100)$
    & $E_{\mathrm{spin}}(D=80)$ & $E_{\mathrm{spin}}(D=100)$ \\ \hline \hline
  22 & 0.139654860817 & 0.139654869116 & 0.807121935054 & 0.807121935023 & -0.434912539817 & -0.434912539817
                                                                             \\ \hline
  42 & 0.086355374176 & 0.086355479294 & 0.775723907194 & 0.775723905709 & -0.438751678570 & -0.438751678582
                                                                             \\ \hline
  62 & 0.064239765116 & 0.064240094937 & 0.760657976140 & 0.760657965104 & -0.440148664874 & -0.440148664972
                                                                             \\ \hline
  82 & 0.051802232654 & 0.051803243348 & 0.751408113523 & 0.751408046065 & -0.440871682329 & -0.440871682728 \\
 \end{tabular}
 \caption{\label{tab:1Supp} Squared staggered magnetization $M_{\mathrm{stag}}^{2}$, mean singlet fraction
          per double well $P_{0}$, and mean spin energy per site $E_{\mathrm{spin}}$, for the AFM of total length
          $N$ obtained from ground state computation with MPS of bond dimension \mbox{$D=80$} and
          \mbox{$D=100$}.}
\end{table}

\begin{table}
 \begin{tabular}{ c || c | c }
  L & $M_{\mathrm{stag}}^{2}(D=80)$ & $M_{\mathrm{stag}}^{2}(D=100)$ \\ \hline \hline
  22 & 0.141157102848 & 0.141157444444 \\ \hline
  42 & 0.088248384680 & 0.088248983897 \\ \hline
  62 & 0.065508480733 & 0.065509337169 \\
 \end{tabular}
 \caption{\label{tab:2Supp} Squared staggered magnetization $M_{\mathrm{stag}}^{2}$ for middle sublattices of length $L=22$, $42$, and $62$, for
                        an AFM of total length $N=82$ obtained from ground state computation with MPS of bond dimension \mbox{$D=80$} and
                        \mbox{$D=100$}.}
\end{table}

\begin{table}
 \begin{tabular}{ c || c | c }
  L & $M_{\mathrm{stag}}^{2}(D=80)$ & $M_{\mathrm{stag}}^{2}(D=100)$ \\ \hline \hline
  22 & 0.141068047422 & 0.141069521513 \\ \hline
  42 & 0.088395582256 & 0.088397878145 \\ \hline
  62 & 0.065999995175 & 0.066003286606 \\
 \end{tabular}
 \caption{\label{tab:3Supp} Squared staggered magnetization $M_{\mathrm{stag}}^{2}$ for middle sublattices of length $L=22$, $42$, and $62$, for
                        an AFM of total length $N=142$ obtained from ground state computation with MPS of bond dimension \mbox{$D=80$} and
                        \mbox{$D=100$}.}
\end{table}

As discussed in the main text, for long enough chains, we observe that 
antiferromagnetic order develops in a middle sublattice faster 
than on the total chain.
We find that the timescales for observables measured on the sublattice
are controlled by the range of the observable itself, as far as finite size effects
can be ignored.
Therefore, in order to quantify this observation, we need to compare the observables in the evolved sublattice with
the corresponding AFM values for a sublattice of the same size in an infinite chain.
The thermodynamic mean singlet fraction $P_{0, \mathrm{TD}} = \ln(2) \approx 0.693$ and mean spin energy $E_{\mathrm{spin}, \mathrm{TD}} = 1/4-\ln(2) \approx -0.44315$ are well known \cite{HulthenSupp}.
The squared staggered magnetization on a finite sublattice, however, cannot be computed exactly, so that we approximate
the reference value by the numerical estimation in a long chain ($N=82$), shown in Tab.~\ref{tab:2Supp}.
Increasing the chain length, the reference values do not change significantly, as one can see in Tab.~\ref{tab:3Supp} for $N=142$.

In 2D, the corresponding values of the AFM were obtained with Quantum Monte Carlo by means of the ALPS code~\cite{ALPSSupp}, and they are listed in Tab.~\ref{tab:4Supp}.

\begin{table}
 \begin{tabular}{ c || c | c }
  N & $M_{\mathrm{stag}}^{2}$ & $E_{\mathrm{spin}}$ \\ \hline \hline
  4 $\times$ 4 & 0.233090090642115 & -0.574325441574560 \\ \hline
  6 $\times$ 6 & 0.164620512963000 & -0.603311944444444 \\ \hline
  8 $\times$ 8 & 0.135035306250000 & -0.618890781250000 \\ \hline
  10 $\times$ 10 & 0.121902426000000 & -0.625150439774003 \\
 \end{tabular}
 \caption{\label{tab:4Supp} Squared staggered magnetization $M_{\mathrm{stag}}^{2}$ and mean spin energy per site
          $E_{\mathrm{spin}}$ for the AFM of total size $N$ obtained from Quantum Monte
          Carlo using the ALPS code~\cite{ALPSSupp}.}
\end{table}

\subsection{Adiabatically evolved states}

In the following, we present the absolute values of our observables for the state obtained at the end of the adiabatic ramping,
in the same sequence as they appear in the main text.

\subsubsection{Absence of holes}

\begin{figure}[h]
\centering
\begin{tabular}{c c c}
\includegraphics[width=0.32\textwidth]{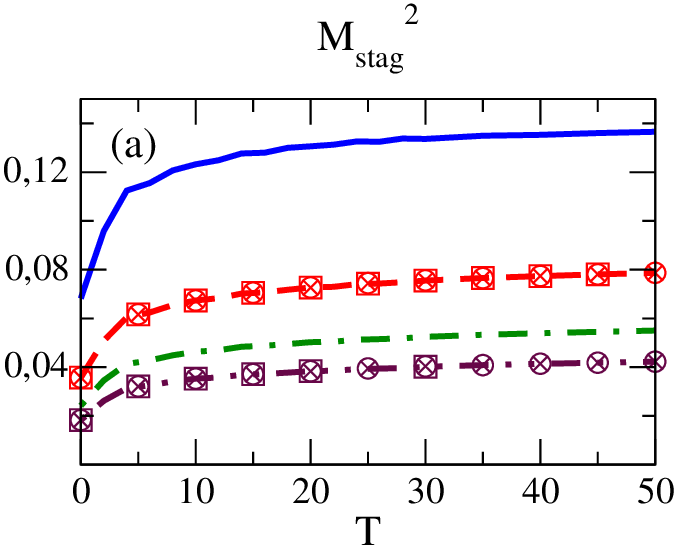} &
\includegraphics[width=0.32\textwidth]{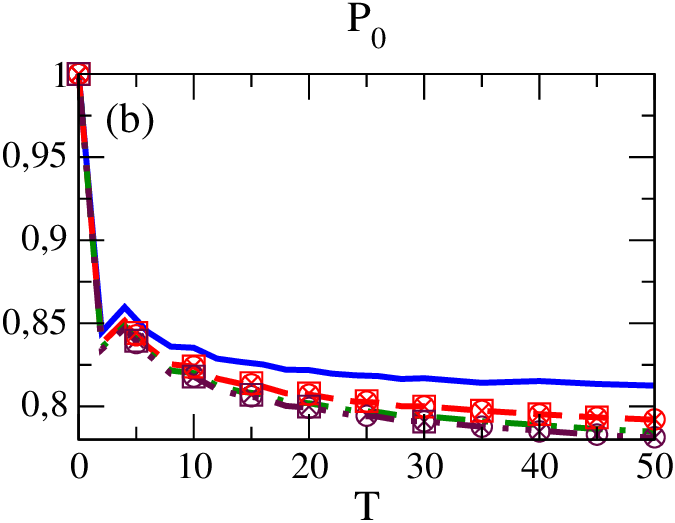} &
\includegraphics[width=0.333\textwidth]{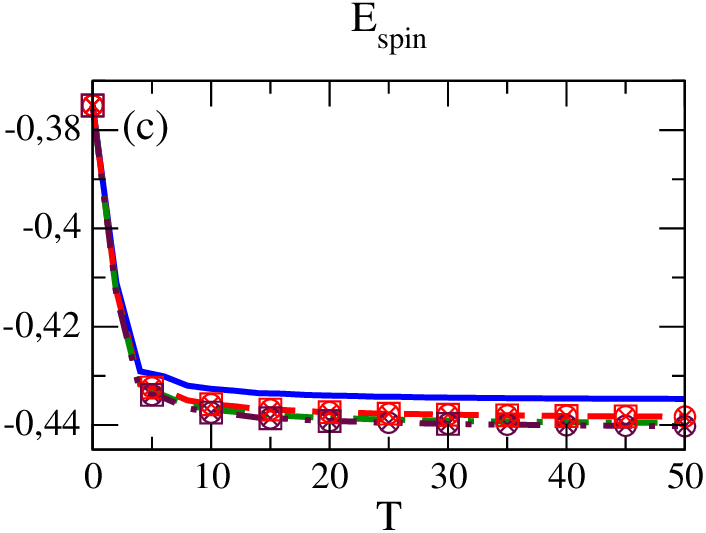}
\end{tabular}
\caption{\label{fig:1Supp} $M_{\mathrm{stag}}^{2}$, $P_{0}$, and $E_{\mathrm{spin}}$, as functions of the ramping time $T$, for chains of length $N=22$ (solid blue), $42$ (dashed red), $62$ (dash-dotted green), and $82$ (dash double-dotted brown).
The results were obtained with MPS of bond dimension $D=60$ and Trotter step $\delta t = 0.02$ (lines), $D=40$ and $\delta t = 0.02$ (circles), $D=40$ and $\delta t = 0.005$ (crosses), and $D=60$ and $\delta t = 0.005$ (squares).}
\end{figure}

\begin{figure}[h]
\centering
\begin{tabular}{c c c}
\includegraphics[width=0.32\textwidth]{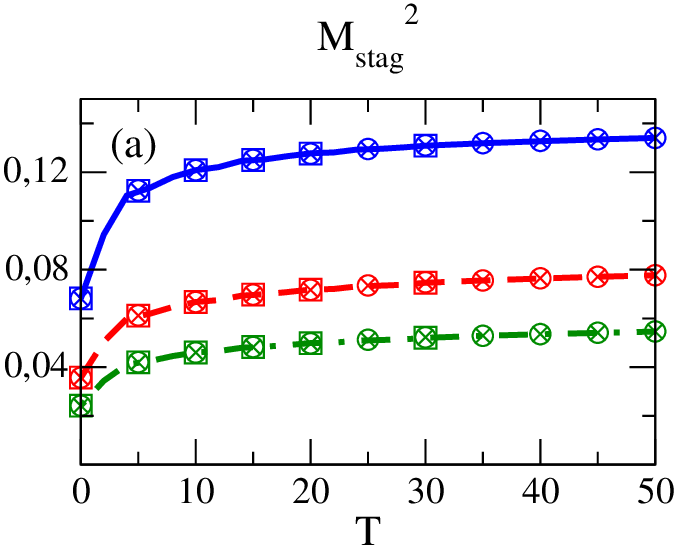} &
\includegraphics[width=0.32\textwidth]{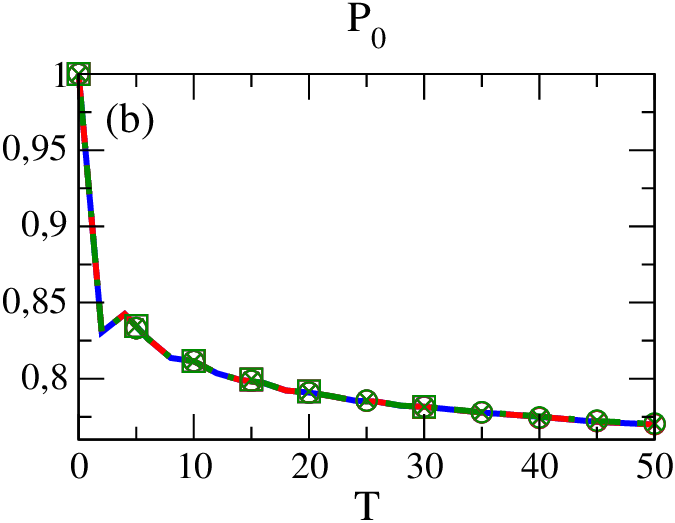} &
\includegraphics[width=0.333\textwidth]{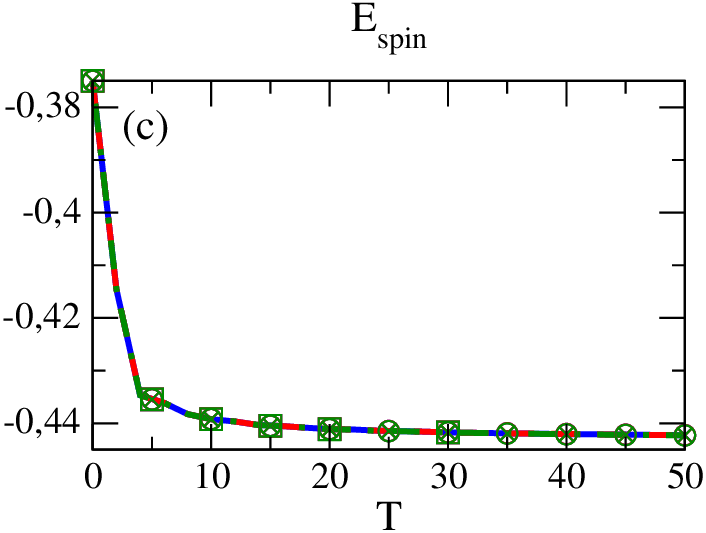}
\end{tabular}
\caption{\label{fig:2Supp} Same quantities as in Fig.~\ref{fig:1Supp}, evaluated on sublattices of length $L=22$ (solid blue), $42$ (dashed red), and $62$ (dash-dotted green), of a total lattice of length $N=82$.}
\end{figure}

Fig.~\ref{fig:1Supp} and \ref{fig:2Supp} show the computed expectation values at the end of the protocol, as a function of the total ramping time, for the case of no holes.
Convergence of the numerical results is checked by comparing the results for bond dimension \mbox{$D=40$} and \mbox{$60$} and for Trotter steps \mbox{$\delta t = 0.02$} and $0.005$. The largest relative error found (for the largest system and the longest ramping time) is of the order of $10^{-4}$,
which ensures enough precision for our analysis.

\subsubsection{Effect of holes}

\begin{figure}[h]
\centering
\begin{tabular}{c c c}
\includegraphics[width=0.32\textwidth]{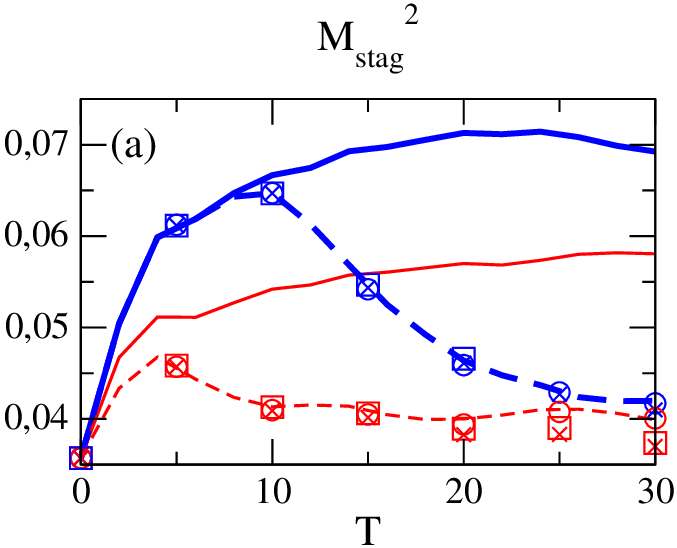} &
\includegraphics[width=0.312\textwidth]{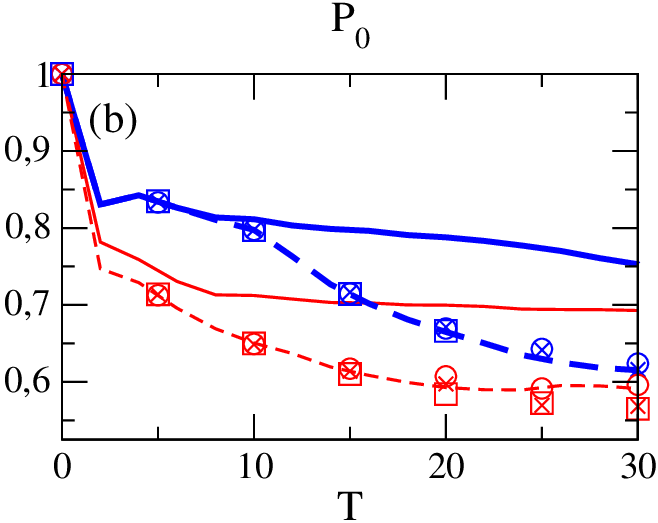} &
\includegraphics[width=0.333\textwidth]{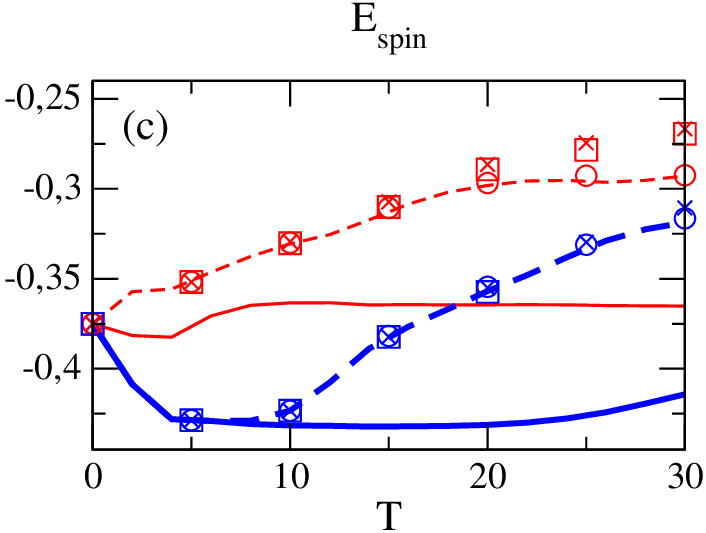}
\end{tabular}
\caption{\label{fig:3Supp} $M_{\mathrm{stag}}^{2}$, $P_{0}$, and $E_{\mathrm{spin}}$, as functions of the ramping time $T$, evaluated on the middle $L=42$ site sublattice, for 2 (solid) and 4 holes (dashed) on $N=82$ sites (thick blue) and $N=42$ sites (thin red), and the holes are initially located at the boundaries, and \mbox{$t=2$}.
The results correspond to $D=60$ and Trotter step $\delta t = 0.02$ (lines), $D=40$ and $\delta t = 0.02$ (circles), $D=40$ and $\delta t = 0.005$ (crosses), and $D=60$ and $\delta t = 0.005$ (squares).}
\end{figure}

Injecting holes into the sample results in a substantial drop of the squared staggered magnetization, and an increase in the energy (Fig.~\ref{fig:3Supp}).
The latter implies that the system gets excited and the numerical simulation via MPS becomes more demanding.
Now, the relative error after the longest ramping time, $T=30$, for the case of 4 holes on $N=82$ sites shown in the main text, is of the order of $0.01-0.1$, but it becomes significantly smaller for shorter times.
This worst-case error does not affect our conclusions, since the main effect we observe, the drop of the magnetization value upon hole arrival, is much larger ($\approx 30\%$) and occurs already at much shorter times ($T \approx 20$), for which the numerical error is only of the order of $10^{-3}$.
The figure also shows that 2 holes on $42$ sites are more dramatic than on $82$ sites, and that the negative effect of holes increases with their number.

\subsubsection{Harmonic trap}

\begin{figure}[h]
\centering
\begin{tabular}{c c c}
\includegraphics[width=0.32\textwidth]{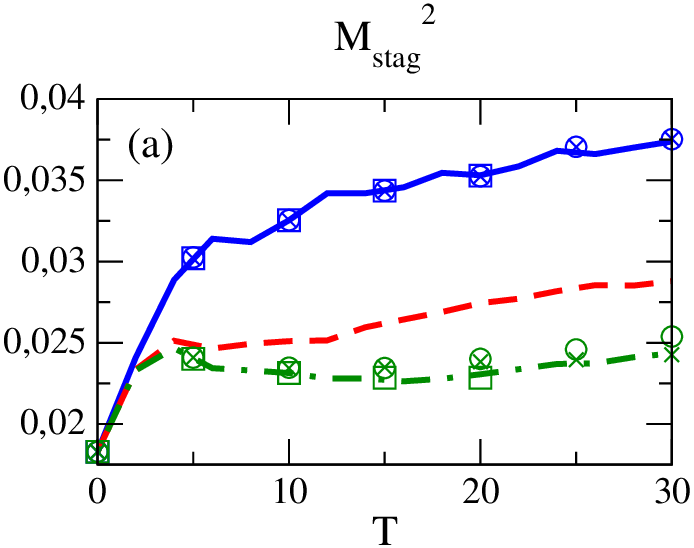} &
\includegraphics[width=0.304\textwidth]{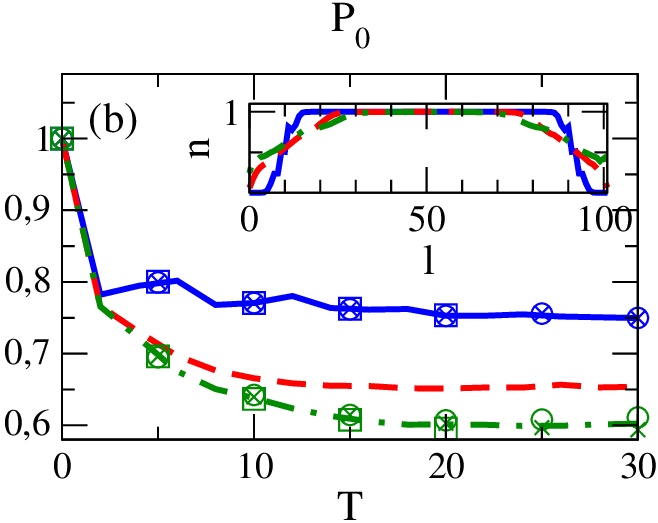} &
\includegraphics[width=0.326\textwidth]{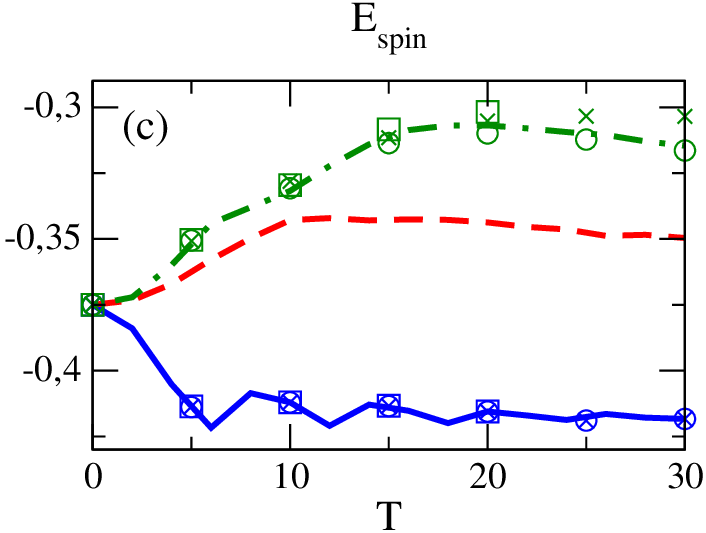}
\end{tabular}
\caption{\label{fig:4Supp} $M_{\mathrm{stag}}^{2}$, $P_{0}$, and $E_{\mathrm{spin}}$, as functions of the ramping time $T$, evaluated on the middle $L=82$ site sublattice, for 10 holes initially on each boundary of $82$ fermions, with a harmonic trap of strength \mbox{$V_{\mathrm{t}}=0.004$} (dash-dotted green), \mbox{$V_{\mathrm{t}}=0.006$} (dashed red), and \mbox{$V_{\mathrm{t}}=0.02$} (solid blue), and \mbox{$t=3$}.
The inset in b) shows the occupation $n$ of lattice site $l$ after ramping time $T=30$, and we find that the holes delocalize precisely $\pm 2t$ at the boundaries of the trap.
Again, the results correspond to $D=60$ and Trotter step $\delta t = 0.02$ (lines), $D=40$ and $\delta t = 0.02$ (circles), $D=40$ and $\delta t = 0.005$ (crosses), and $D=60$ and $\delta t = 0.005$ (squares).}
\end{figure}

By including a harmonic trap \mbox{$V_{\mathrm{t}} \sum_{k} (k - k_{0})^{2} \hat{n}_{k}$}, the holes can be confined outside of the sample.
We consider 10 holes left and 10 holes right of a sample of size $82$, and successively increase the trap strength, as shown in Fig.~\ref{fig:4Supp}.
Consistent with the results in the previous section, our simulation is most demanding for the weakest trap, when holes can still enter the sample and excite the system.
The largest relative error in that case is of the order of $0.01-0.1$, but it decreases significantly if the trap strength is increased.
Again, this worst-case error does not affect any of our conclusions.

\subsection{Two dimensional case}

\begin{figure}[h]
\centering
\begin{tabular}{c c c}
\includegraphics[width=0.32\textwidth]{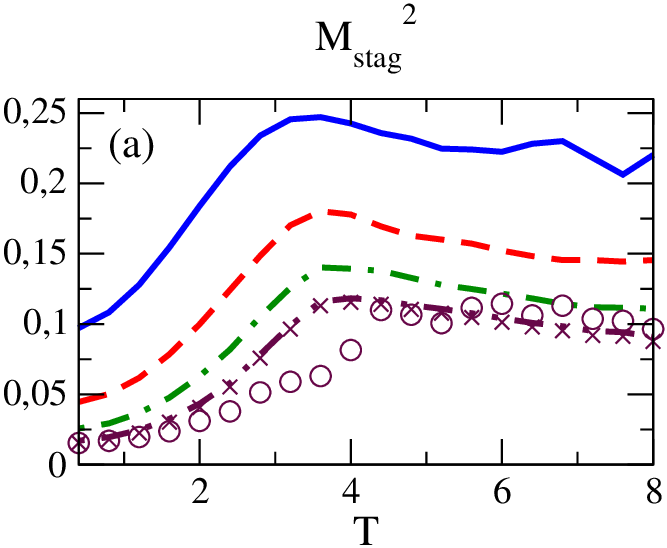} &
\includegraphics[width=0.314\textwidth]{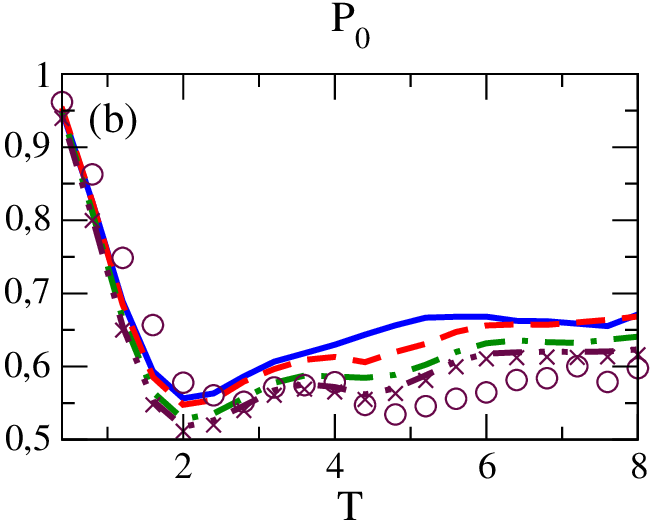} &
\includegraphics[width=0.327\textwidth]{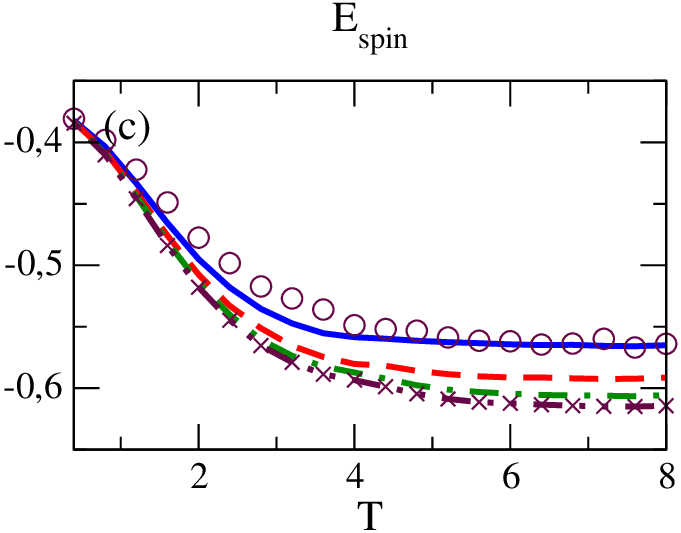}
\end{tabular}
\caption{\label{fig:5Supp} $M_{\mathrm{stag}}^{2}$, $P_{0}$, and $E_{\mathrm{spin}}$, as functions of the ramping time $T$, for $N=4 \times 4$ (solid blue), $6 \times 6$ (dashed red), $8 \times 8$ (dash-dotted green), and $10 \times 10$ (dash double-dotted brown).
The results were obtained with PEPS of bond dimension $D=4$ and Trotter step $\delta t = 0.03$ (lines), $D=2$ (circles), and $D=3$ (crosses).}
\end{figure}

\begin{figure}[h]
\centering
\begin{tabular}{c c}
\includegraphics[width=0.32\textwidth]{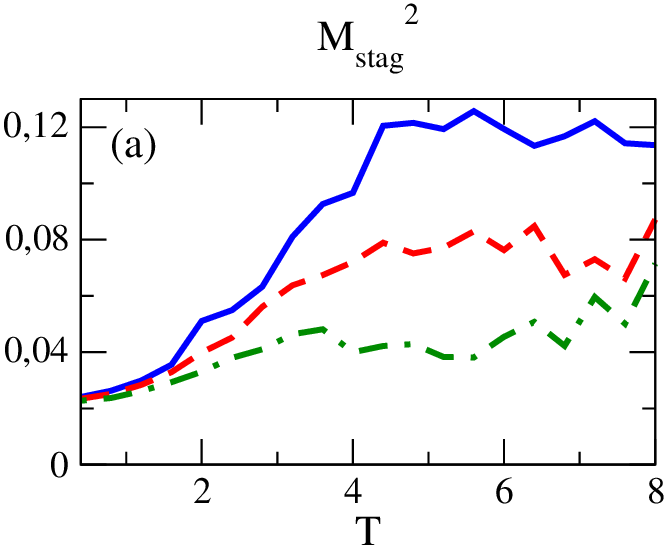} &
\includegraphics[width=0.322\textwidth]{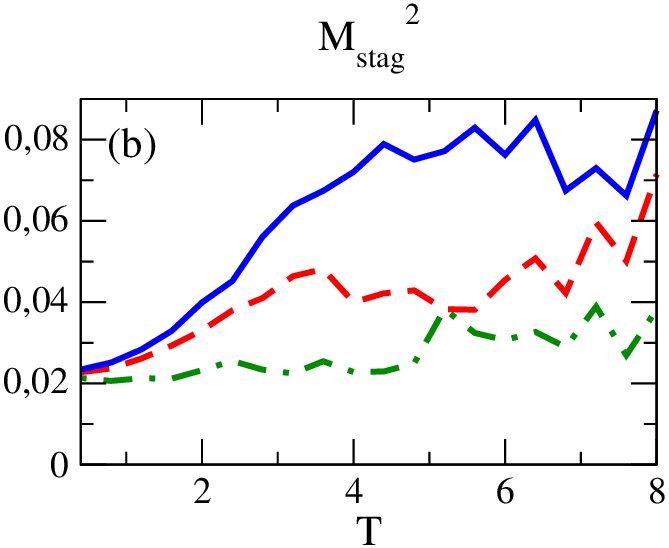}
\end{tabular}
\caption{\label{fig:6Supp}(a) $M_{\mathrm{stag}}^{2}$ as a function of the ramping time $T$, for $N=8 \times 8$ without holes (solid blue), with 1 hole (dashed red), and 2 holes (dash-dotted green), where the holes are initially localized at the boundary, and $t=2.5$.
(b) $N=8 \times 8$ with 4 holes initially distributed at the boundary, with no harmonic trap (dash-dotted green), and with a trap of strength
\mbox{$V_{\mathrm{t}}=0.25$} (dashed red), and \mbox{$V_{\mathrm{t}}=2.5$} (solid blue), and again \mbox{$t=2.5$}.
The results correspond to $D=2$ and Trotter step $\delta t = 0.03$.}
\end{figure}

In 2D, the time evolution is done with PEPS of bond dimension \mbox{$D=2$}, \mbox{$D=3$} and \mbox{$D=4$}, and the largest relative error is of the order of $10^{-2}$, for the largest system in Fig.~\ref{fig:5Supp}.
Our results suggest that qualitative insight can already be gained from PEPS with $D=2$.
Therefore, Fig.~\ref{fig:6Supp} shows the effect of holes and harmonic trap for $D=2$ without convergence check.
Just as in 1D [Fig.~\ref{fig:3Supp}(a)], the staggered magnetization decreases significantly with increasing number of holes, and the harmonic trap confines the holes on the outside.

\begin{thebibliography}{99}

\bibitem{Esslinger} T. Esslinger, Annu.\ Rev.\ Condens.\ Matter Phys.\ {\bf 1}, 129 (2010).

\bibitem{Joerdens1} R. J\"ordens {\it et al.}, \nat {\bf 455}, 204 (2008).

\bibitem{Schneider1} U. Schneider {\it et al.}, \sci {\bf 322}, 1520 (2008).

\bibitem{Werner} F. Werner {\it et al.}, \prl {\bf 95}, 056401 (2005).

\bibitem{Joerdens2} R. J\"ordens {\it et al.}, \prl {\bf 104}, 180401 (2010).

\bibitem{McKay} D. C. McKay and B. DeMarco, Rep.\ Prog.\ Phys.\ {\bf 74}, 054401 (2011).

\bibitem{Ho} T.-L. Ho, arXiv:0808.2677.

\bibitem{Sorensen} A. S. S\o{}rensen {\it et al.}, \pra {\bf 81}, 061603(R) (2010).

\bibitem{Schneider2} U. Schneider {\it et al.}, \emph{Supporting Online Material}, \sci {\bf 322}, 1520 (2008).

\bibitem{Schneider3} U. Schneider {\it et al.}, arXiv:1005.3545v1.

\bibitem{MPS} G. Vidal, \prl {\bf 93}, 040502 (2004) (we use the method introduced in: F. Verstraete, J. J. Garc\'ia-Ripoll and J. I. Cirac, \prl {\bf 93}, 207204 (2004)).

\bibitem{PEPS1} F. Verstraete and J. I. Cirac, arXiv:cond-mat/0407066v1.

\bibitem{PEPS2} V. Murg, F. Verstraete and J. I. Cirac, \pra {\bf 75}, 033605 (2007).

\bibitem{Sebby-Strabley} J. Sebby-Strabley {\it et al.}, \pra {\bf 73}, 033605 (2006).

\bibitem{Trotzky1} S. Trotzky {\it et al.}, \sci {\bf 319}, 295 (2008).

\bibitem{Trotzky2} S. Trotzky {\it et al.}, \prl {\bf 105}, 265303 (2010).

\bibitem{Jaksch} D. Jaksch {\it et al.}, \prl {\bf 81}, 3108 (1998).

\bibitem{Ramping} Other ramping schemes \mbox{$J_{o}(\tau) = J(\tau/T)^{x}$} for different values of $x$ give qualitatively similar results.

\bibitem{Zener} C. Zener, Proc.\ R.\ Soc.\ Lond.\ A {\bf 137}, 696 (1932).

\bibitem{Matsumoto} M. Matsumoto {\it et al.}, \prb {\bf 65}, 014407 (2001).

\bibitem{Bruun} G. M. Bruun {\it et al.}, \pra {\bf 80}, 033622 (2009).

\bibitem{Hulthen} L. Hulth\'en, Ark.\ Mat.\ Astron.\ Fys.\ {\bf 26A}, 1 (1938).

\bibitem{Sandvik} A. W. Sandvik and H. G. Evertz, \prb {\bf 82}, 024407 (2010).

\bibitem{Supplementary} See Supplemental Material for details to our numerical method and errors, and absolute values of experimental observables.

\bibitem{Bond} MPS and PEPS are characterized by a bond dimension $D$ that determines their number of variational parameters, and convergence of numerical results to the exact solution is checked by increasing $D$. The time evolution is done with a second (1D) and first (2D) order Trotter decomposition of the evolution operator, and hereby convergence is checked by decreasing the Trotter step $\delta t$~\cite{Supplementary}.

\bibitem{Affleck} I. Affleck {\it et al.}, J.\ Phys.\ A: Math.\ Gen.\ {\bf 22}, 511 (1989).

\bibitem{InPreparation} M. Lubasch {\it et al.}, (to be published).

\bibitem{Boninsegni} M. Boninsegni and N. V. Prokof'ev, \prb {\bf 77}, 092502 (2008).

\bibitem{BosonicAFM} The hardcore bosonic system has a ferromagnetic interaction, but the antiferromagnet can be obtained as the highest excited state \cite{Sorensen, Garcia-Ripoll}.

\bibitem{Garcia-Ripoll} J. J. Garc\'ia-Ripoll, M. A. Martin-Delgado and J. I. Cirac, \prl {\bf 93}, 250405 (2004).

\end{thebibliography}

\begin{thebibliography}{99}

\bibitem{WhiteDMRGSupp} S. R. White, \prl {\bf 69}, 2863 (1992);
                        U. Schollw\"ock, \rmp {\bf 77}, 259 (2005).

\bibitem{DMRGSupp} G. Vidal, \prl {\bf 93}, 040502 (2004).

\bibitem{MPS1Supp} F. Verstraete, D. Porras and J. I. Cirac, \prl {\bf 93}, 227205 (2004).

\bibitem{MPS2Supp} F. Verstraete, J. J. Garc\'ia-Ripoll and J. I. Cirac, \prl {\bf 93}, 207204 (2004).

\bibitem{PEPS1Supp} F. Verstraete and J. I. Cirac, arXiv:cond-mat/0407066v1.

\bibitem{PEPS2Supp} V. Murg, F. Verstraete and J. I. Cirac, \pra {\bf 75}, 033605 (2007).

\bibitem{GSmpsSupp} F. Verstraete and J. I. Cirac, \prb {\bf 73}, 094423 (2006);
                    M. B. Hastings, J. Stat. Mech. 2007, P08024 (2007).

\bibitem{GSpepsSupp} M. B. Hastings, \prb {\bf 73}, 085115 (2006).

\bibitem{PEPS3Supp} F. Verstraete, V. Murg and J. I. Cirac, Adv.\ Phys.\ {\bf 57}, 143 (2008).

\bibitem{HulthenSupp} L. Hulth\'en, Ark.\ Mat.\ Astron.\ Fys.\ {\bf 26A}, 1 (1938).

\bibitem{ALPSSupp} F. Albuquerque {\it et al.}, Journal of Magnetism and Magnetic Materials {\bf 310}, 1187 (2007).

\end{thebibliography}
\end{document}